\documentstyle[epsf,here,psfig,11pt]{article}
\begin {document}
\title{Towards Solving QCD in Light-Cone Quantization --  \break
       On the Spectrum of the Transverse Zero Modes for SU(2).}
\author{Hans-Christian Pauli and Rolf Bayer \\ 
        Max-Planck-Institut f\"ur Kernphysik \\
        D-69029 Heidelberg \\ }
\date{30 September 1995}
\maketitle

\begin{abstract}
The formalism for a non-abelian pure gauge theory in (2+1) dimensions 
has recently been derived within Discretized Light-Cone Quantization,  
restricting to the lowest {\it transverse} momentum gluons.    
It is argued why this model can be a paradigm for full QCD. 
The physical vacuum becomes non-trivial even in light-cone 
quantization. The approach is brought here to tractable form by 
suppressing by hand both the dynamical gauge and the constraint zero
mode,  
and by performing a Tamm-Dancoff type Fock-space truncation. 
Within that model the Hamiltonian is diagonalized numerically,
yielding  
mass spectra and wavefunctions of the glue-ball states. We find that 
only color singlets have a stable and discrete bound state spectrum.
The connection with confinement is discussed. The structure function 
of the gluons has a shape like $ [{x(1-x)}] ^{1\over 3} $. 
The existence of the continuum limit is verified by deriving 
a coupled set of integral equations.
\end{abstract} 
\vfill
\noindent 
Preprint MPI H-V32-1995 (Revised edition) 

\newpage 
\section{Introduction and Motivation}
Constructing even the lowest state that is the `vacuum' of a 
Quantum Field Theory has been so notoriously difficult that the 
conventional Hamiltonian approach was given up altogether, 
long ago in the Fifties.
It was overlooked that other forms, particularly Dirac's `front form' 
of Hamiltonian dynamics \cite{Dir49} might have less severe problems.
In fact, concretely realizing the front form with periodic boundary 
conditions
one might combine the aspects of a simple vacuum \cite{Wei66} with a 
careful treatment of the infrared degrees of freedom. 
The method referred to is 
Discretized Light-Cone Quantization (DLCQ) \cite{PaB85a}. 
A review can be found in \cite{PaB91}.
Potentially, the method is able to reconcile the simplistic
but otherwise successful constituent quark picture of Feynman's 
parton model \cite{Fey69} with the low-energy regime of Quantum 
Chromodynamics (QCD) as the fundamental theory of hadrons.
But the conversion of this non-perturbative method into a reliable
tool  
for hadronic physics is bestowed with many difficulties \cite{Gla95}, 
one of them being the so-called `zero mode problem'. 

Recently, (2+1) dimensional pure SU(2) gauge theory has been examined  
suppressing transverse gluon momentum excitations \cite{PKP95}.
The dimensionally-reduced theory turns out to be SU(2) gauge theory 
in (1+1) dimensions coupled to adjoint scalar matter.
A topological (gauge) mode appears coupled to true dynamical 
Fock modes of the transverse gluon fields.
Moreover, a constrained zero mode appears which is defined by 
a linear but still very complicated equation, whose inversion is far 
from being trivial \cite{Kal95}, structures that 
were foreseen by Franke et al. \cite{FNP81} in (3+1) dimensions. 
Such types of models, first discussed in DLCQ 
by \cite{DaK93,DKB93,BDK93,AnD95}  
but without zero modes and assuming only color singlet string states, 
enable insight into how to overcome the obstacles in the full theory.

Next to interesting vacuum structures \cite{Kal95}, 
the dimensionally reduced model of QCD has a fascinating excitation 
spectrum whose structure is difficult to guess, see \cite{PKP95}. 
It is this latter point to which we address ourselves in the present 
work. We do that at the expense of rigor, keeping only the Fock space 
structure of the model to be presented below in detail.  
Despite the heavy truncations the present work has rather interesting  
aspects, which can serve as paradigms for what one expects in full
QCD. 

\section{Formulation of the Model}

The model \cite{PKP95} considered in the sequel is a (1+1) dimensional
non-abelian gauge theory covariantly coupled to scalar adjoint matter
whose Lagrangian density is given by 
\begin {equation}
    {\cal L} = {\rm Tr} \Bigl(
        - {1\over2} {\bf F}^{\alpha\beta} {\bf F}_{\alpha \beta} 
       + {\bf D}^\alpha \Phi   {\bf D}_\alpha \Phi 
       - \mu  _0 ^2 \Phi \Phi \Bigr) 
\,. \end   {equation}
The color electro-magnetic fields are the usual ones in matrix
notation, {\it i.e.} 
\begin {equation} 
    {\bf F}^{\alpha\beta} \equiv \partial^\alpha  {\bf A}^\beta  
   - \partial^\beta  {\bf A}^\alpha 
   + i g \bigl[ {\bf A}^\alpha, {\bf A}^\beta\bigr] \ 
 \equiv \partial^\alpha  {\bf A}^\beta  
  - {\bf D}^\beta  {\bf A}^\alpha 
\,. \label {eq:co-emag-fields} \end {equation} 
The covariant derivative $ {\bf D}_\alpha$ is the carrier of 
`longitudinal gauge invariance' which is not affected 
by adding a mass term. 
Fixing the gauge
\begin {equation} 
    {\rm Tr} \left( \tau ^\pm {\bf A} ^+ \right) = 0
    \,,\qquad\partial _- {\bf A} ^+ = 0
    \,,\qquad{\rm and}\quad
    {\rm Tr} \left< \tau ^3   {\bf A} ^- \right> _{0} = 0
\,, \label {eq:gauge} \end {equation} 
completes the model.
With the purpose of setting the notation we recall here in short 
the most important steps of Ref.\cite{PKP95}.

For SU(2), the above $\tau ^a$ are 2$\times$2 matrices related 
to the Pauli spin matrices $\sigma _i$. 
They define a color helicity basis in terms of which all 
fields are expanded, {\it i.e.} 
\begin {equation}
  \tau ^3 = {1\over2} \sigma _z \ , \quad 
  \tau ^\pm \equiv {1\over{2\sqrt{2}}} (\sigma _x \pm i \sigma _y)
\ , \qquad {\rm thus} \quad 
  \Phi = \tau ^3 \varphi _3 + \tau ^+ \varphi_+ + \tau ^- \varphi_- 
\ . \end    {equation}
We shall be careful to have the Lorentz and color indices 
consistently raised and lowered, respectively, when we 
write down indiviual field components.
Since the matrices are traceless and hermitean, one deals 
with nine real-valued operator functions, three for each field.
Gauge fixing, {i.e.} $  {\bf A} ^+  = v \tau ^3 $,
reduces the problem to six  real-valued operator functions and 
one quantum mechanical operator $ v \equiv v  (x^+)$, which one 
keeps track of in the combination $z\equiv vgL/\pi$.
Four out of these seven have a non-vanishing conjugate momentum,
particularly the momenta canonically conjugate to 
$ v $, $ \varphi _3$, $ \varphi _-$, and $ \varphi _+$.
Only these are therefore `independent fields' subject to be quantized.
The vector potential $ v $ (thus $z$) is quantized like a quantum 
mechanical variable \cite{Man85} that is like 
$ \bigl[ z , p _z \bigr] = i $, with the momentum
$ p _z \equiv 2\pi \partial _+ v/ g $ conjugate to $z$.
With $ \varphi _+ = \varphi _-^\dagger $ and imposing periodic
boundary conditions (DLCQ), the remaining two fields 
are quantized canonically. 
It is justified \cite {PKP95} to restrict to the 
{\it fundamental modular domain} 
($ 0 < z < 1 $), in which they are represented by
\begin {eqnarray} 
  \varphi_3 (x^-) & = & {a_0 \over \sqrt{4\pi}} 
   + {1 \over \sqrt{4\pi}} \sum _{n =1} ^\infty 
  \Bigl(\  a_n \  w_n \ {\rm e}^{-i n   {\pi\over L}   x^-}
  + \  a^\dagger_n \  w_n \ {\rm e}^{+i n {\pi\over L} x^-} \Bigr) 
\ , \label {phi3exp} \\ 
  {\rm and} \quad \varphi_- (x^-) & = & 
  {{\rm e}^{+ i {\pi\over 2 L}   x^-} \over \sqrt{4\pi}}
  \ \sum_{m={1\over 2}} ^\infty \Bigl(
 \  b_m \  u_m \ {\rm e}^{-i  m   {\pi\over L}   x^-}
 + \ d^\dagger_m \ v_m \ {\rm e}^{+i m {\pi\over L} x^-} \Bigr) 
\ . \label {fockexp} \end    {eqnarray} 
The single particle operators $a_n$, $b_m$ and $d_m$ are the 
carriers of the operator structure and obey 
conventional commutator relations like 
$ \bigl[ a _n , a^\dagger _{n^\prime} \bigr] = \delta_{n,{n^\prime}} $
and 
$ \bigl[ b _m , b^\dagger _{m^\prime} \bigr] = 
  \bigl[ d _m , d^\dagger _{m^\prime} \bigr] = \delta_{m,{m^\prime}}$
with all others vanishing except those involving the zero mode $a_0$. 
The coefficients    
\begin {equation}
    w_n  = 1/ \sqrt{n} \ , \quad
    u_m  = 1/\sqrt{m + \zeta} \ , \quad {\rm and} \quad
    v_m  = 1/\sqrt{m - \zeta} 
    \ , \qquad{\rm with}\quad  \zeta \equiv  z - { 1\over 2} 
\ , \label {coefficients} \end    {equation} 
are real and depend on the vector potential $ v $ through  $\zeta$. 
It is justified to think of the 
`a'-, `b'-, and `d'-particles as of photons, electrons, and positrons,
except that they are all bosons. Below, use of this will be made 
both in the notation and the diagrammatical representation.

Neither $a_0$, the zero mode of $\varphi_3$, nor the remaining three
fields  
$ A _3 ^- $, $ A _- ^- $ and $ A _+ ^- $ have a conjugate momentum. 
They cannot be quantized canonically but determine themselves by 
certain equations of motion. 
The $ A _a ^- $ are determined by the three Gauss' equations. 
In terms of the current 
\begin {equation}
   {\bf J} ^\beta = {1\over i} 
         \bigl[ \Phi , {\bf D} ^\beta \Phi \bigr] 
\label {eq:current} \end {equation}
and its density components 
$ J_ a ^+ = 2 {\rm Tr} \left(\tau ^a {\bf J} ^+ \right)$
they read as
\begin {equation}
  -\partial_-^2  A _3 ^- =  g  J _3 ^+ \ , \quad
  -(\partial_- + i g   v)^2  A _+ ^- = g  J _+ ^+ 
  \ , \quad {\rm and} \quad
  -(\partial_- - i g   v)^2  A _+ ^- = g  J _- ^+ 
\ . \label {gausscomp} \end   {equation}
The first of them can be solved only if the zero mode \cite{PKP95} 
of the r.h.s 
\begin {equation}
   \left\langle J _3 ^+ \right\rangle _{\! 0} 
   \equiv {1\over 2L} \int _{-L} ^{L} d x ^- J _3 ^+ (x^-) 
\,, \qquad{\rm thus}\quad 
   Q _3 = 2L \left\langle J _3 ^+ \right\rangle _{\! 0} 
   = \sum_{m = {1\over 2}}^{\infty}   
     \Bigl( d^\dagger_m d_m - b^\dagger_m b_m \Bigr)
\,, \label {gaussop} \end    {equation}
vanishes.  
This cannot be satisfied as an operator, but must be used to  
select out physical states. To find them is easy: 
they  must have the same total number of `b' and `d' particles. 
When discussing below the spectra we will see, that this condition is
different from the naive picture of color singlets.
Finally, one must express $ a_0$ in terms of the independent fields.
The defining {\it constraint equation} is linear in $ a _0$, 
for details see Ref.\cite{PKP95}.
In the solution, $a_0$ should be an explicit functional 
of the Fock-space operators, 
and a function of $\zeta$ and $\mu$, {\it i.e.}
\begin {equation}
   a_0 \equiv a_0 \left[ a_n ^\dagger, b_m ^\dagger, d_m ^\dagger
                       , a_n , b_m , d_m ; \zeta , \mu \right]
\,. \label {constraint_solution} \end {equation}
Thus far, only approximate solutions have been constructed 
\cite {Kal95}.

\section{On the Hamiltonian and its Diagonalization}
\label {sec:energy_model}

Upon evaluation \cite{PKP95}, the total (light-cone) momentum $ P ^+$ 
becomes a diagonal and thus simple Fock-space operator 
\begin {equation}
   P ^+ =   
     {\pi\over L } \sum_{n =1} ^\infty n \  a^\dagger_n   a_n  
   + {\pi\over L } \sum_{m = {1\over2}} ^\infty \Bigl[ 
      (m+\zeta)\  b^\dagger_m  b_m 
    + (m-\zeta)\  d^\dagger_m  d_m \Bigr]
\,. \label {mommat} \end   {equation}
The total (light-cone) energy $ P ^-$, the `Hamiltonian' of the
theory, is a off-diagonal Fock-space operator \cite{PKP95}. 
It is so complicated that we only outline its construction 
and refer for details to \cite{PKP95}. 
The defining equation is with $\hat g \equiv g/\sqrt {16\pi}$
\begin {eqnarray} 
     P ^- 
 &=& - 4 \hat g ^2 {L \over \pi} { d^2 \over dz^2}
  +  {1 \over 2} \int \limits _{- L}^{+ L} \!\! dx^-   \Bigl( 
  \mu _0^2\,( \phi _3 \phi _3 + \phi _+ \phi _- + \phi _- \phi _+ )
  +   \partial_-  A ^- _3 \partial_-  A ^- _3 \Bigr) 
\nonumber \\ 
 &+&  {1 \over 2} \int \limits _{- L}^{+ L} \!\! dx^- \Bigl( 
      (\partial_- +i g  v)  A ^- _+
      (\partial_- -i g  v)  A ^- _-
    + (\partial_- -i g  v)  A ^- _-
      (\partial_- +i g  v)  A ^- _+             \Bigr) 
\,. \label {eq:hamiltonian} \end {eqnarray} 
To get it as a Fock-space operator one expresses the
charge densities of Eq.(\ref {eq:current}) in terms of the
$\phi _a$ and substitutes Eqs. (\ref {phi3exp}) and (\ref {fockexp}).
Inverting Eqs.(\ref {gausscomp}) yields $ A ^- _a$ in terms of the
Fock-space operators and of $a _0$, thus \cite{PKP95}
\begin {equation}
   P ^- = P ^- \left[ a_0; a_n ^\dagger, b_m ^\dagger, d_m ^\dagger
                       , a_n , b_m , d_m ; \zeta , \mu \right]
\,. \label {eq:Hamil_a0} \end {equation}
Finally, one has to invert the constraint equation for $a_0$
and to substitute Eq.(\ref {constraint_solution}) everywhere. 
As a results one gets the Hamiltonian in terms of 
`raw-ordered' Fock-space operator products.
It is reasonable to rewrite these 
`time ordered products' as normal ordered products 
plus the sum of all pairwise contractions.
The totally contracted terms are identical with the Fock-space 
vacuum expectation value which we conveniently define as
$ W(\zeta) \equiv 
 \left\langle 0 \right| P ^- \left| 0  \right\rangle $.
Contrary to conventional DLCQ these may not be discarded since 
as functions of $\zeta$ they are part of the operator structure. 
Terms with one creation and one destruction operator are usually 
referred to as the `contraction terms' or the 
`self-induced inertias' $ P ^- _C$, 
those with two creation and two destruction operators 
as `seagulls' $P ^- _S$, those with one creation and three 
destruction operators as `forks' $P ^- _F$. 
Explicit formulas can be found in Appendices~\ref {sec:contractions}, 
\ref {sec:seagulls}, and \ref {sec:forks}, respectively.
They have been derived for $a_0=0$, but one should emphasize that
a non-vanishing $a_0$  contributes to the 
`gauge potential energy' $W(\zeta)$ as well as to the contractions, 
seagulls and forks. In general, the Hamiltonian becomes thus 
\begin {equation}
     P ^- = - 4 \hat g ^2 {L \over \pi} 
            { d^2 \over d\zeta^2} + W(\zeta) 
          + P ^- _C (\zeta) + P ^- _S (\zeta) + P ^- _F (\zeta) 
\,. \label {exact_hamiltonian} \end {equation}
In principle one has to diagonalize simultaneously
momentum, energy and charge, {\it i.e.} 
\begin {equation} 
    P ^+    \left| {\Psi_i} \right\rangle 
  = P ^+ _i \left| {\Psi_i} \right\rangle  
\ , \quad 
    P ^-    \left| {\Psi_i}  \right\rangle 
  = P ^- _i \left| {\Psi_i}  \right\rangle 
\ , \quad {\rm and} \qquad
    Q _3    \left| {\Psi_i}  \right\rangle = 0
\ , \label {eigenvalue_eqn} \end {equation} 
repectively.
The denumerable eigenvalues of momentum and energy are denoted by
$P ^+ _i$ and $P ^- _i$, respectively. 
For the momentum and the charge this is not difficult, 
since any arbitrary Fock state 
\begin {equation}
   \left| \nu  \right\rangle 
            = a_{n_1} ^\dagger a_{n_2} ^\dagger \dots
              b_{m_1} ^\dagger b_{m_2} ^\dagger \dots
              d_{p_1} ^\dagger d_{p_2} ^\dagger \dots 
   \left| 0 \right\rangle 
\end {equation}
with the same number of b- and d-particles is a solution, 
labelled appropriately by $ \nu = 1,2,\dots, N$.
The number of states $N$ is finite for any fixed value of the 
`harmonic resolution' \cite{PaB85a} $K \equiv L P^+ _i / \pi$.
Since energy and momentum commute, they span a complete set 
of states for diagonalizing the energy. 
The Fock space vacuum $\left| 0  \right\rangle $, defined by
\begin {equation}
    a_n \left| 0   \right\rangle = 0 
\ , b_m \left| 0   \right\rangle = 0 \ , \quad{\rm and}\quad
   d_m \left| 0   \right\rangle = 0 
\ , \end {equation}
has no particle content and thus vanishing momentum. 

Next to being an operator in Fock space the Hamiltonian is a 
Schr\"odinger operator with respect to the gauge field $v$ or $\zeta$.
Generating a complete set of functions $\psi _n (\zeta)$ by solving 
a Schr\"odinger equation with a convenient single particle 
potential $V(\zeta)$, 
\begin {equation}
   \Bigl( - 4 \hat g ^2 { d^2 \over d\zeta^2} 
   + V(\zeta) \Bigr) \psi_n (\zeta)
   = \omega _n \psi_n (\zeta)
\ , \label {gauge_potential} \end {equation}
the solutions to  Eq.(\ref {eigenvalue_eqn}) 
{\it must have a non-separable structure}, {\it i.e.}
\begin {equation}
  \left| \Psi _i  \right\rangle = \sum _{n=0} ^\infty \sum _{\nu=1} ^N
   C ^{(i)} _{n,\nu} \ \psi _n (\zeta) \left| \nu  \right\rangle 
   , \quad{\rm with}\qquad 
   C ^{(i)} _{n,\nu} \not= c ^{(i)} _n c ^{(i)} _\nu
\ . \end {equation}
This statement holds in general, since the coefficient functions
of the Fock-space operators in the Hamiltonian depend on $\zeta$
in a complicated way. 
The physical vacuum $\left| {vac}  \right\rangle $
that is the state with lowest energy of the full theory will thus
acquire structure with respect to $\zeta$ as in \cite {PKP95}, 
see also \cite{Kal95}, 
entirely due to a `physical gauge' like $\partial _- {\bf A} ^+ = 0 $ 
as opposed to the `light-cone gauge' $ {\bf A} ^+ = 0 $. 
But these structures will be disregarded in the rest of the paper 
where the vacuum is simple due to the model assumptions.

\section{The Model within the Model}
\label {sec:formulate_problem}

In the present work we approach the problem from the tail. 
We are interested in the particle sector, its spectral properties
and wavefunctions. As a first step, in order to face a tractable
problem, 
we do {\it not neglect} but {\it omit by hand} $a_0$, and in addition 
suppress the fluctuations in the gauge mode, {\it i.e.}
\begin {equation}
   a_0 \equiv 0 
   \quad{\rm and} \qquad 
   \psi _n (\zeta) \equiv \delta _{n,0} \, \delta(\zeta) 
\ , \end {equation}
see also Ref.\cite{PKP95}. 
For $\zeta\equiv 0$ the Hamiltonian becomes then
\begin {equation}
   P ^- = P ^- _{\rm Fock} \equiv 
   P ^- _C (0) + P ^- _S (0) + P ^- _F (0) 
\,. \label {model_hamiltonian} \end {equation}
One should emphasize that this model does {\it not correspond} 
to the light-cone gauge since we keep $gvL = \pi/2$ as the static 
value around which the dynamic treatment would fluctuate.
Last not least, we perform a Tamm-Dancoff
truncation by restricting to the lowest, the 2-particle sectors.
This `model within the model' will provide useful insight into the 
structure of the full solution.

The `contraction terms' $ P ^- _C$ as given in 
Appendix~\ref {sec:contractions} should deserve some explanatory 
remarks. In a first step one gets for them rather straightforwardly
\begin {equation}
   {\pi \over L} P^- _C 
  = \sum\limits_{n=1}^\infty 
    {I _n (\zeta) \over n} a_n^\dagger a_n 
  + \sum\limits_{m=\frac{1}{2}}^\infty
    {J _m (+\zeta) \over m+\zeta } b_m^\dagger b_m 
  + \sum\limits_{m=\frac{1}{2}}^\infty
    {J _m (-\zeta) \over m-\zeta } d_m^\dagger d_m 
\ . \end {equation}
The `self-induced inertias' or `tadpole diagrams' $I$ and $J$ contain 
logarithmically diverging pieces,  
\begin{eqnarray}
   I _n (\zeta) 
   = \mu _0 ^2 
   + \hat g ^2 \!\sum\limits_{m=\frac{1}{2}}^\infty
     \Bigg[ \hspace*{-0.5cm} & & 
      {2\over {\left( m + \zeta \right) }} + 
      {4n\over {{{\left( m - n + \zeta  \right) }^2}}} - 
      {4n\over {{{\left( m + n + \zeta  \right) }^2}}} 
\nonumber \\ 
  &\hspace*{-.25cm}+\hspace*{-.25cm}&  \left. 
      {2\over {\left( m - \zeta  \right) }} 
    + {4n\over {{{\left( m - n - \zeta  \right) }^2}}} 
    - {4n\over {{{\left( m + n - \zeta  \right) }^2}}} \right]
\ , \label {self_ind_I} \\ 
     J _m (\zeta) 
   = \mu _0 ^2 
   + \hat g ^2 \sum\limits_{n=1}^\infty
   \Bigg[ \hspace*{-0.5cm} & & 
   {2\over {n}} + 
   {4\left( m + \zeta \right) 
     \over {{{\left( m - n + \zeta  \right) }^2}}} - 
   {4\left( m + \zeta \right) 
     \over {{{\left( m + n + \zeta  \right) }^2}}}\Bigg]
\nonumber \\ 
   + \hat g ^2 \sum\limits_{p=\frac{1}{2}}^\infty 
  \Bigg[ \hspace*{-0.5cm} & & 
   {1\over {\left( p - \zeta \right) }} + 
   {1\over {\left( p + \zeta \right) }} + 
   {4\left( m + \zeta \right) \over {{{\left( m - p \right) }^2}}} - 
   {4\left( m + \zeta \right) \over {{{\left( m + p \right) }^2}}}  
\Bigg] , \label {self_ind_J} \end {eqnarray}
through terms like $\sum _{n=1} ^\infty 1/n$.
They can be cancelled by renormalizing the mass like 
\begin {equation}
    \mu_0^2 
  = \mu ^2 - 16 \hat g ^ 2
  - 2 \hat g ^2 \left( \sum_{m={1\over2}} ^\infty \frac{1}{m} 
  + \sum_{n=1} ^\infty \frac{1}{n} \right) 
  = \mu ^2   - (16 + C) \hat g ^ 2
  - 4 \hat g ^2 \sum_{m=1} ^\infty \frac{1}{n} 
\ , \label {def:renormalization} \end {equation}
with $\hat g = g/\sqrt {16\pi} $. 
Important is that the renormalized mass $\mu$ is {\it finite}.
Its value is not unique, and thus both the `16' and the constant 
\begin {equation}
   C \equiv  2\sum_{m={1\over 2}} ^\infty \frac {1} {m} 
           - 2\sum_{n=1}          ^\infty \frac {1} {n} 
     = 4 \ln {2}  
\label {def:constant_c} \end {equation}
could be absorbed into the renormalized $\mu$. But here is the
problem:  
The sums in Eqs. (\ref {self_ind_I}) and (\ref {self_ind_J}) 
are not completely symmetric in the integers and half-integers.
Converting the expressions tabulated in 
Table~\ref {tab:contractions} generates the constant $C$
in either one place or the other. In the continuum limit
the two sums in Eq.(\ref {def:constant_c}) tend to cancel in line 
with Section~\ref {sec:integral_equation}.
We therefore shall work with $C=0$ whenever not mentioned otherwise.

\section{Numerical Solutions}
\label {sec:numerical_solution}
Consider two types of 2-particle Fock states, referred to as the 
`aa'- and the `bd'-space, respectively, 
\begin {eqnarray}
   \left| {n}  \right\rangle _a 
&=& a_n ^\dagger a_{K-n} ^\dagger \left| 0  \right\rangle 
   \quad{\rm for }\quad  n = 1,\dots, \left[ \frac{K}{2} \right]
\ , \nonumber \\ 
   \qquad \mbox{and}\qquad 
   \left| {m}  \right\rangle _b 
&=& b_m ^\dagger d_{K-m} ^\dagger \left| 0 \right\rangle 
   \quad{\rm for }\quad  m = \frac{1}{2} ,\dots, K - \frac{1}{2} 
\ . \label {def:space} \end {eqnarray}
They are orthogonal, $\left\langle m \vert n \right\rangle = 0$, 
and simultaneous eigenstates of $Q_3$ and $P ^+$, {\it i.e.}
\begin {eqnarray}
  Q_3  \left| {n}  \right\rangle _a  =  0     \quad &,& \quad
  P ^+ \left| {n}  \right\rangle _a  
  =  {\pi \over L} K \left| {n}  \right\rangle _a 
\ , \\ 
  Q_3  \left| {m}  \right\rangle _b  =  0     \quad &,& \quad
  P ^+ \left| {m}  \right\rangle _b  
  =  {\pi \over L} K \left| {m}  \right\rangle _b 
\ , \end {eqnarray}
for all values of $n$ and $m$, 
as they should according to Eq.(\ref {eigenvalue_eqn}).
Gauss' equation (\ref {gausscomp}) prevents us from including also
`charged' states like for example 
$ b_m ^\dagger a_{K-m} ^\dagger \left| 0  \right\rangle $.
Any of the $K + [K/2] $ linear superposition of these basis states 
like
\begin {equation}
   \left| {\Psi_i} \right\rangle 
   = \sum _{n=1} ^{[{K \over 2}]} 
     \left\langle {n} \right| \widetilde C _a \,
     \left| {i}  \right\rangle \ \left| {n} \right\rangle _a 
   + \sum _{m={1\over 2}} ^{K - {1 \over 2}} 
     \left\langle {m} \right| \widetilde C _b 
     \left| {i}  \right\rangle \ \left| {m} \right\rangle _b 
   = \sum _{n=1} ^{[{K \over 2}]} 
     \left\langle {n} | \Psi _{i,a}\right\rangle\, 
     \left| {n}  \right\rangle _a 
   + \sum _{m={1\over 2}} ^{K - {1 \over 2}} 
     \left\langle {m} | \Psi _{i,b}\right\rangle\, 
     \left| {m} \right\rangle _b 
\label {def:eigenfunctions} \end {equation}
are thus  eigenfunctions of $Q _3$ and $P ^+$ with the same
eigenvalues, 
while the unitary matrix $U =  \widetilde C _a +  \widetilde C _b$ 
determines itself by diagonalizing $P ^-$, with the ($K + [K/2]$) 
eigenvalues $P ^- _i$ and subject to normalization
\begin {equation}
    \left\langle {i} \right| U ^\dagger U \left| {i}  \right\rangle =
      1 
    \,, \qquad{\rm for\ all} \ i
\,. \label {def:normalization} \end {equation}
Since $ P ^\nu P_\nu = 2 P ^+ P ^-$ is to be interpreted as the
operator 
of invariant mass squared $ M^2 $, 
the eigenvalues of $P ^-$ will be presented below as the product 
$ 2 P ^+_i P ^- _i \equiv M _i ^2$. As unit of mass we shall use 
$ m_u= \hat g = g/\sqrt{16\pi}$.
Note that the coupling constant $g$ has here the dimension of a mass.

As a rather welcome advantage of a DLCQ calculation, the 
diagonalization of $100\times 100$ matrices takes only a couple of 
milliseconds on a modern work station. One thus could go to fairly
large values of the resolution within our simple model. 
However, in order to unravel the structure of the spectra 
and not to be overwhelmed by a flood of data, 
we first keep the mass parameter fixed to $\mu = 0 $
and restrict ourselves to $ 20 \leq K \leq 80 $. 
In view of typical values for the harmonic
resolution $K\leq 22$ quoted in the 
literature \cite {DaK93,DKB93,BDK93,AnD95}  
the above range is still large for practical purposes.

\begin{figure}[t]
\vspace*{-4cm}
\centerline{
\psfig{figure=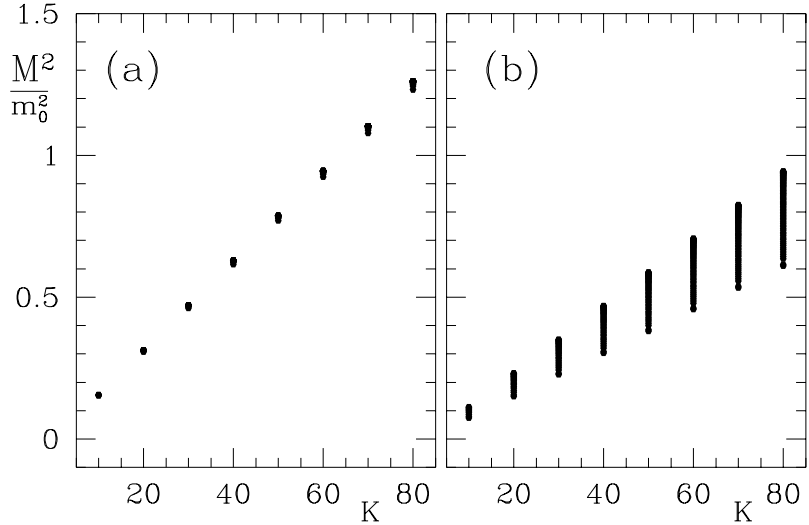,height=12cm,rheight=8cm,angle=-0}
}\vspace*{4cm}
\centerline{ \parbox{35em}{
\caption{ \sl \label {aabd} 
 Invariant mass squared eigenvalues $M ^2 _i$ 
 {\it versus} Harmonic Resolution $K$.~-- 
 Left side (a): The Fock space is restricted to {\it only} the 
 aa-states $\left| {n}  \right\rangle _a$. 
 Note the almost complete degeneracy
 of all $K/2$ states.      
 Right side (b): The Fock space is restricted to {\it only} the 
 bd-states $\left| {n}  \right\rangle _b$.~--
 Note: All eigenvalues increase roughly linear in $K$, and show no 
 trend to stabilize.~--
 Parameters values are $\mu^2=0$ and $m_0 ^2 = 2 (100 m_u) ^2 $. 
} } }
\end{figure}

\underline{1. The reduced 2-particle Fock spaces.}
When one restricts to {\it only} the aa-space, the Hamiltonian matrix
$ \left\langle {n^\prime} \right| P^- \left| {n}  \right\rangle $ 
is diagonal from the outset, 
since the a-particles have no interaction according to 
Table~\ref {tab:seagulls}. They only have self-induced inertias.
To first order of approximation, for moderately small values of $\mu$
these inertias are linear in the momentum. The eigenvalues
of $ P ^- $ become approximately independent of $K$ and thus 
the invariant masses therefore linear in $K$ and highly 
degenerate, as displayed in Figure~\ref {aabd}a as function of $K$. 

If one restricts oneself to {\it only} the bd-space, 
the Hamiltonian matrix is non-diagonal due to the interaction
between the b- and the d-particles.
As displayed in Figure~\ref {aabd}b, this causes 
the eigenvalues to spread over a wider range of energies but 
obviously not strong enough so as to stabilize the low energy
parts of the spectrum. One understands this behaviour analytically, 
see Section~\ref {sec:integral_equation}.

\underline{2. The 2-particle Fock space.}
The behaviour changes drastically if one includes simultaneously the 
aa- and the bd-states. One now allows for the virtual scattering 
from the aa-space into the bd-space and back, 
represented by the matrix elements
$S_{aa} (n_1, n_2; n_3, n_4)$ in Table~\ref {tab:seagulls}.
\begin{figure}[t]
\vspace*{-2cm}
\centerline{
\psfig{figure=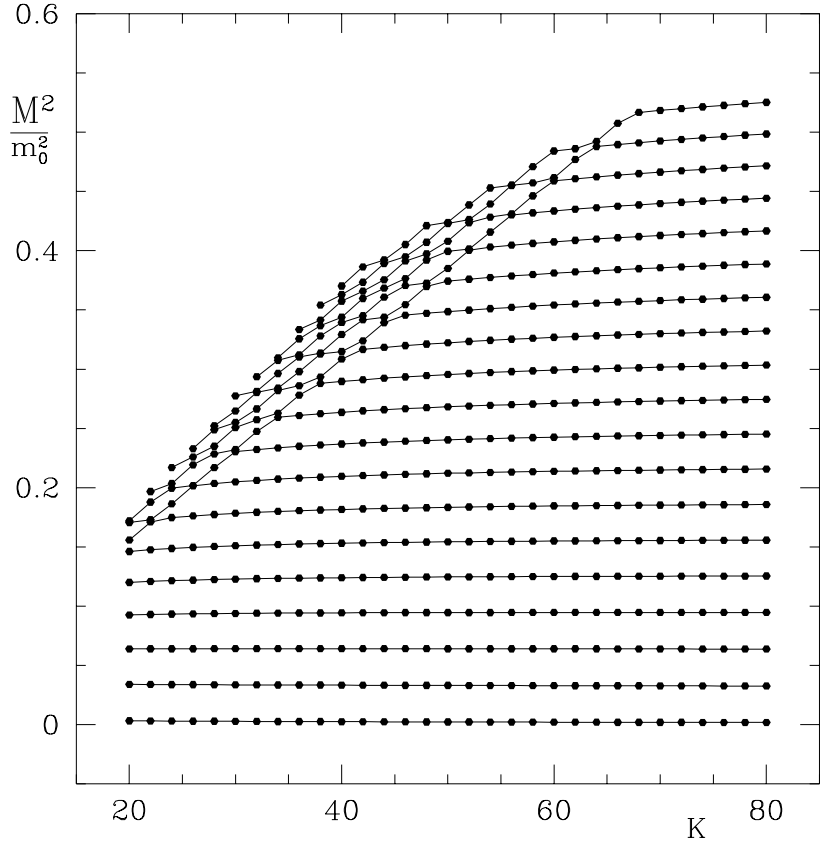,height=12cm,rheight=8cm,angle=-0}
}\vspace*{4cm}
\centerline{ \parbox{35em}{
\caption{ \label {spectrum1} {\sl
 Invariant mass squared eigenvalues $M ^2 _i$ 
 {\it versus} Harmonic Resolution $K$.~-- 
 The Fock space includes both the aa-  and bd-states. 
 Note: The lowest eigenvalues $M_i ^2$ are roughly independent of $K$,
 with a roughly constant spacing.~-- 
 Parameters values are $\mu^2=0$ and $m_0 ^2 = 2 (100 m_u) ^2 $. 
}}}}
\end{figure}
In Figure~\ref {spectrum1}, the first 19 mass squared eigenvalues 
are plotted as functions of $K$. At a first glance the eigenvalues
are $K$-independent, with a roughly equal spacing like the states 
in a harmonic oscillator well. 
At the left upper corner of the figure seem to develop some
`crossings' of the states, which we shall resolve and explain below.
At a second glance, the spacing is not completely independent
of the energy, rather it decreases slowly with increasing energy, 
typically like the eigenstates in a potential with linearly 
increasing walls.

We like to emphasize the discreteness of the spectrum.
Discreteness surviving the continuum limit
$K \rightarrow \infty$ should be considered as the earmark of 
confinement, even if such a statement seems premature within 
the present 1+1 dimensional `model within a model'. 
Most if not all field theories particularly QED 
confine in 1+1 dimension \cite{epb87,epk95}.
Remarkable however is that one gets such a spectrum at all, 
in a model with no other ingredients than pure gauge theory.

A word of caution seems in order. The restriction to the 2-particle 
sector is {\it ad hoc} and should not be overemphasized. Including
more and more particles, one would expect structures which
would tend to true continua for ever increasing $K$.
At least parts of the so obtained spectra would correspond to 
`physical glue-balls in relative motion'.
In short and over-stressing the point, one would expect spectra 
which qualitatively resemble the multi-particle spectra of QED 
in (1+1) dimensions, as displayed in the figures of 
Ref.\cite {epb87,epk95}.
\begin{figure}[t]
\vspace*{-2cm}
\centerline{
\psfig{figure=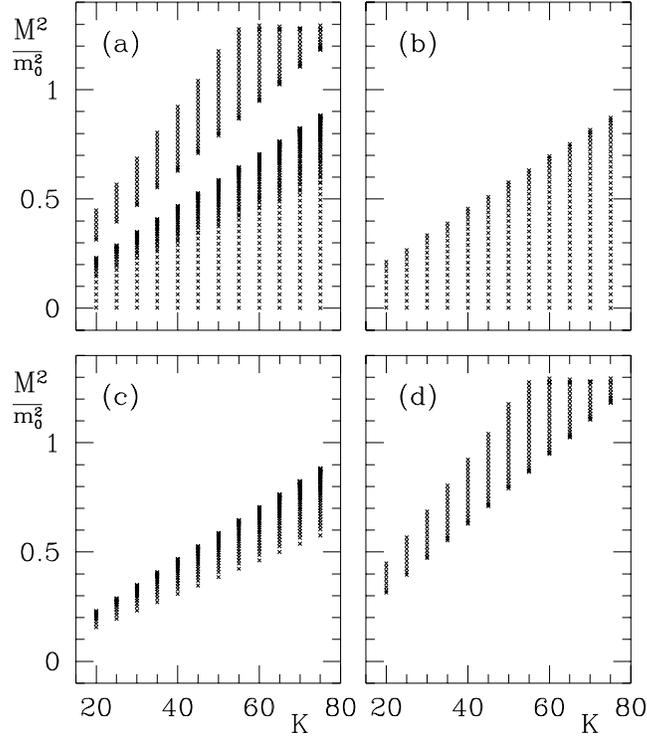,height=11cm,rheight=8cm,angle=-0}
}\vspace*{2.5cm}
\centerline{ \parbox{35em}{
\caption{ \label {separate} {\sl
 Spectral Decomposition with Multiplet $Q$.~-- 
  For each state $\left| {i,K}  \right\rangle $ 
  the multiplet expectation value $Q$ 
  is calculated, and the spectrum disentangled correspondingly, 
  see text.~-- 
 Upper  left (a): Full spectrum.
 Upper right (b): Spectrum of the quasi-singlets   ($ Q = 0$). 
 Lower  left (c): Spectrum of the quasi-triplets   ($ Q = 1$). 
 Lower right (d): Spectrum of the quasi-pentuplets ($ Q = 2$).--
 Note: Only the singlet states have a stable spectrum.~--
 Parameters values are $\mu^2=0$ and $m_0 ^2 = 2 (100 m_u) ^2 $. 
}}}}
\end{figure}

\underline{3. Symmetries and Multiplet structure.}
In the naive thinking of color invariance a color singlet should 
have a finite spectrum. Are our eigenstates color singlets?
Opposed to this interpretation is that the spectrum diverges
when restricting to either aa- or bd-states {\it alone}
as displayed in Figure~\ref {aabd}.
Only when both are included, one has a chance for having wave 
functions which are invariant under rotations in color space. 
Can one be more specific? What are the {\it symmetries} of our
problem? 

By inspecting the numerical results for the eigenfunctions one
observes three distinctly different classes: 
(I)   those with $ \Psi _{i,a} \sim \Psi _{i,b} $, 
(II)  those with $ \Psi _{i,a} \sim 0 $, and 
(III) those with $ \Psi _{i,a} \sim -2 \Psi _{i,b} $,
for comparable size of the single particle momenta.
A separation into two groups can be based on an exact symmetry:
the Hamiltonian is invariant when exchanging the 
b- and the d-particles. Denoting the corresponding charge-conjugation 
operator by $C _\pi$, its eigenvalues must be $C _\pi ^\prime = \pm
1$. 
In fact {\it all} numerical eigenfunctions are {\it either}
charge-conjugation even {\it or} odd, by inspection, and the 
latter coincides with the class II states.
Aiming at a measure to classify the states we introduce the operators
\begin {equation}
   Q _- = \sum _{n=1} ^\infty 
          a _n ^\dagger b_{n+{1\over2}} 
        - a _n          d_{n-{1\over2}} ^\dagger 
   \ , \qquad
   Q _+ = \sum _{n=1} ^\infty 
          a _n          b_{n+{1\over2}} ^\dagger 
        - a _n ^\dagger d_{n-{1\over2}} 
\,. \label {def:Q_+} \end {equation}
Together with $ Q _3 $ they satisfy SU(2) commutation relations in the
Weyl representation
\begin {equation}
   \left[ Q _3 , Q _- \right] = Q _-     \ , \qquad
   \left[ Q _+ , Q _3 \right] = Q _+     \ , \qquad
   \left[ Q _- , Q _+ \right] = Q _3 
\,, \end {equation}
with the group invariant
\begin {equation}
    Q ^2 \equiv Q _3 Q_3 + Q _- Q_+ + Q _+ Q_-
\ . \end {equation}
The eigenfunctions of the Hamiltonian cannot be 
eigenfunctions to $Q^2$ since the two do not commute, see also
Section~\ref {sec:charges} below.
One can however calculate the {\it expectation values} 
$ q_i \equiv \left\langle {\Psi_i} \right| Q^2 
  \left| {\Psi_i}  \right\rangle $. 
The `effective eigenvalues' $Q_e$, defined by $Q_e(Q_e+1) = q_i$, 
turn out to be close to the numbers $0,1,2,\dots$. 
We therefore define
\begin {equation}
    Q = \cases{ 0, &if $ 0.0 < Q_e < 0.3$ (class I); \cr
                1, &if $ 0.7 < Q_e < 1.3$ (class II); \cr
                2, &if $ 1.7 < Q_e < 2.3$ (class III). \cr     }
\label {def:Q} \end {equation}
It is remarkable, that none of the expectation values drops out of
the comparably narrow limits set in these equations. Is this a 
consequence of a residual symmetry in the Hamiltonian?~--
For 2 particles the largest possible value is $Q=2$.
As it turns out, all numerical eigenstates have a charge-conjugation 
parity $C _\pi ^\prime = (-1) ^Q$, particularly the 
triplet is charge-conjugation odd.
In Figure~\ref {separate} the same spectrum as in Figure~\ref
{spectrum1}  
is displayed, but separated according to the multiplet-value $Q$.
As it turns out, the crossing situation mentioned in the context 
of Figure~\ref {spectrum1} is due to the crossing between $Q=0$ 
and $Q=1$ states. {\it All} eigenvalues of the singlet $Q=0$ become
virtually {\it independent} of the resolution. As opposed to this
{\it all} members of the triplets or pentuplets, $Q=1$ or $Q=2$, 
respectively, are at least linear in $K$, tending to infinity
in the continuum limit. 

The latter result is remarkable since it is one of the
few direct evidences that only singlets can have finite
masses in a non-abelian theory. 
Some subtleties related to this interpretation will be discussed
further below in Section~\ref {sec:charges}.
\begin{figure}[t]
\vspace*{-2cm}
\centerline{
\psfig{figure=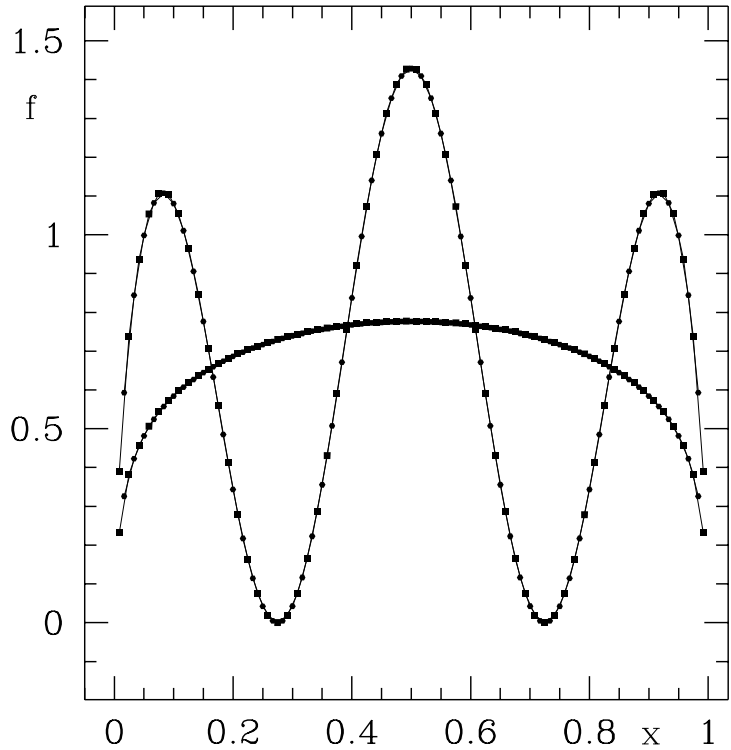,height=11cm,rheight=8cm,angle=-0}
}\vspace*{3cm}
\centerline{ \parbox{35em}{
\caption{ \label {structure1} {\sl
 Structure Functions {\it versus} Momentum Fraction $x$.~-- The structure 
 function is calculated both for the a- and the b-particles, 
 ($x \equiv n/K$, squares) and ($x \equiv m/K$, bullets), respectively, 
 for the ground and the first excited state.~--
 Note: The curve interpolating between squares and bullets is remarkably 
 smooth. Note also, that the structure function of the excited state 
 has more nodes than the one of the ground state, 
 in accord with expectation.~--
 Parameters values are $\mu^2=0$ and $K=60$.
}}}}
\end{figure}

\underline{4. Structure Functions.}
Next to the spectrum we investigate explicitly the eigenfunctions.
Due to diagonalization one knows them for all eigenstates, 
particularly their projections on the Hilbert space,
$ \left\langle n | \Psi_i\right\rangle 
= \left\langle {n} \right| \widetilde C _a 
  \left| {i}  \right\rangle $
and 
$ \left\langle m | \Psi_i\right\rangle 
= \left\langle {m} \right| \widetilde C _b 
  \left| {i}  \right\rangle $, 
see also Eq.(\ref {def:eigenfunctions}).
The probabilities to find an a- or a b-particle with 
longitudinal momentum
$ p ^+ = n {\pi\over L} $ or $ p ^+ = m {\pi\over L} $ are given by
\begin {equation} 
   \left\langle {\Psi_i} \right| a _n ^\dagger a _n 
   \left| {\Psi_i}  \right\rangle 
   = \left\langle {n} \right| \widetilde C _a 
     \left| {i}  \right\rangle ^2
\ , \ {\rm and}\qquad 
   \left\langle {\Psi_i} \right| b _m ^\dagger b _m 
   \left| {\Psi_i}  \right\rangle 
   = \left\langle {m} \right| \widetilde C _b 
     \left| {i}  \right\rangle ^2
\ , \end {equation} 
repectively, and are related to the structure functions $f (x)$ by
\begin {equation} 
   f _i^{(a)} ( n/K )  
   = K \left\langle {\Psi_i} \right| a _n ^\dagger a _n 
       \left| {\Psi_i}  \right\rangle 
\ , \ {\rm and}\qquad 
   f _i^{(b)} ( m/K )  
   = K \left\langle {\Psi_i} \right| b _m ^\dagger b _m 
       \left| {\Psi_i}  \right\rangle 
\ . \end {equation} 
In the continuum, the probability to find a parton with 
longitudinal momentum fraction between $x=n/k$ and $x+dx$
is $dx \, f (x)$. Note that the structure functions are {\it not} 
normalized to unity, because of Eq.(\ref {def:normalization}), 
rather they obey the {\it sum rule}
$ 
  \int _0 ^1 dx \left( 
   {1\over 2} f _i^{(a)} (x) + f _i^{(b)} (x) \right) = 1
$. 
 In displaying them in Figure~\ref {structure1}, we restrict ourselves
to the first two states. Due to the residual SU(2) symmetry the
structure functions for the a- and b-particles turn out to be 
extremely similar, for which reason they have been compiled in the
same plot. By inspection the structure function for the first state
can be approximated by
\begin {equation} 
   f_\alpha (x) = {2\over 3} 
            {\Gamma(2+2\alpha) \over 
       \left[\Gamma(1+ \alpha) \right] ^2} 
       \left[ \, x(1-x) \, \right] ^\alpha
\ , \label {strufu}\end {equation} 
with an exponent numerically closer to $\alpha \sim {1\over 3}$
than to $\alpha \sim {1\over 4}$.
All higher states have more nodal structures.
\begin{figure}[t]
\vspace*{-2cm}
\centerline{
\psfig{figure=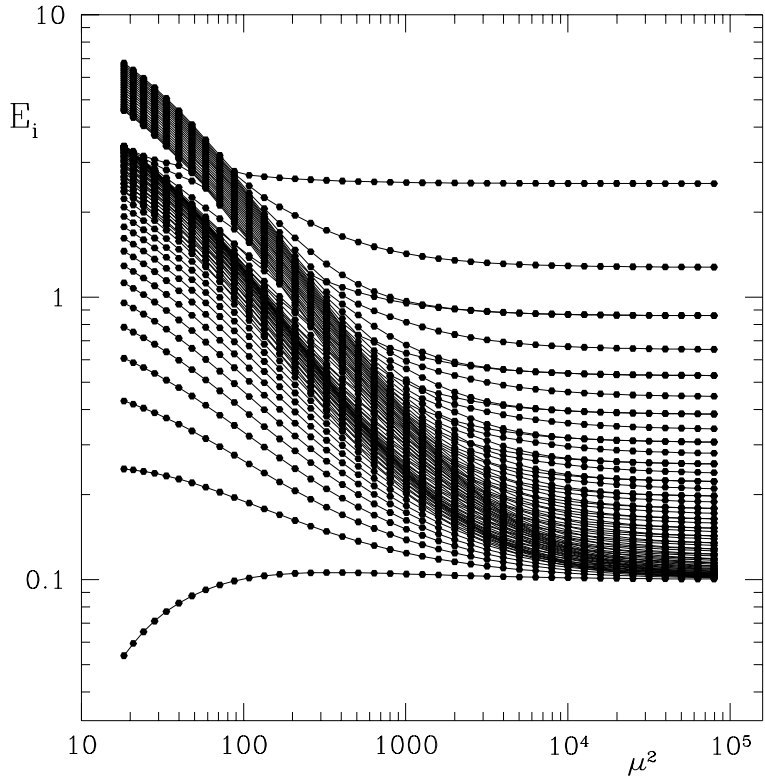,height=11cm,rheight=8cm,angle=-0}
}\vspace*{3cm}
\centerline{ \parbox{35em}{
\caption{ \label {mudependence1} {\sl
 Invariant mass squared eigenvalues $E_i=2M_i^2 / (1000m^2_u+40\mu^2)$ 
 {\it versus} Mass Parameter $\mu ^2$. 
 Note the logarithmic scale and how the spectrum changes from being 
 `interaction dominated' at smaller to being
 `mass dominated' at larger values of $\mu$.~--
 The harmonic resolution is fixed at $K = 50$.
}}}}
\end{figure}

\underline{5. Mass-Dependence.}
The results presented thus far have been calculated for 
a vanishing value of the mass parameter $\mu$. 
For a mass parameter $\mu$ much larger than the effective
coupling constant $\hat g$ one can omit the interaction
to first order of approximation. With increasing $\mu$, 
the spectrum must become more and more like the one of two free 
massive bosons, that is like $\sim {\mu^2 \over x(1-x)}$. 
Correspondingly, the structure function of the lowest
state will be peaked at $x=1/2$ 
with an ever decreasing width like a $\delta$-function.
As shown in Figure~\ref {mudependence1}, this is exactly what 
happens. Keeping $\hat g$ fixed and increasing $\mu$ 
the spectrum changes from being `interaction dominated' at
$\mu \sim 0$ in the manner as having been shown in 
Figure~\ref {spectrum1} to being `mass dominated' beyond 
$\mu \sim 100 m_u$. A good earmark of the latter is the
decreasing level density with increasing excitation.

\underline{6. Does the Continuum Limit exist?}
The spectra shown in Figs.~\ref {spectrum1} and~\ref {separate} 
for $ 20 \leq K \leq 80$ did appear to be stable 
as function of $K$. Are they really? What happens when one increases
$K$ by order of magnitudes? Does the continuum limit exist?~--
The question can be answered either analytically by converting the 
Hamiltonian matrix equation to an integral equation, 
or numerically by increasing the {\it order of magnitude} of $K$.
We have done both. The continuum limit is discussed in detail
in Section~\ref {sec:integral_equation}. Here we present the
numerical results, by varying $K$ up to $K \sim 1000$. 
\begin{figure}[t]
\vspace*{-2cm}
\centerline{
\psfig{figure=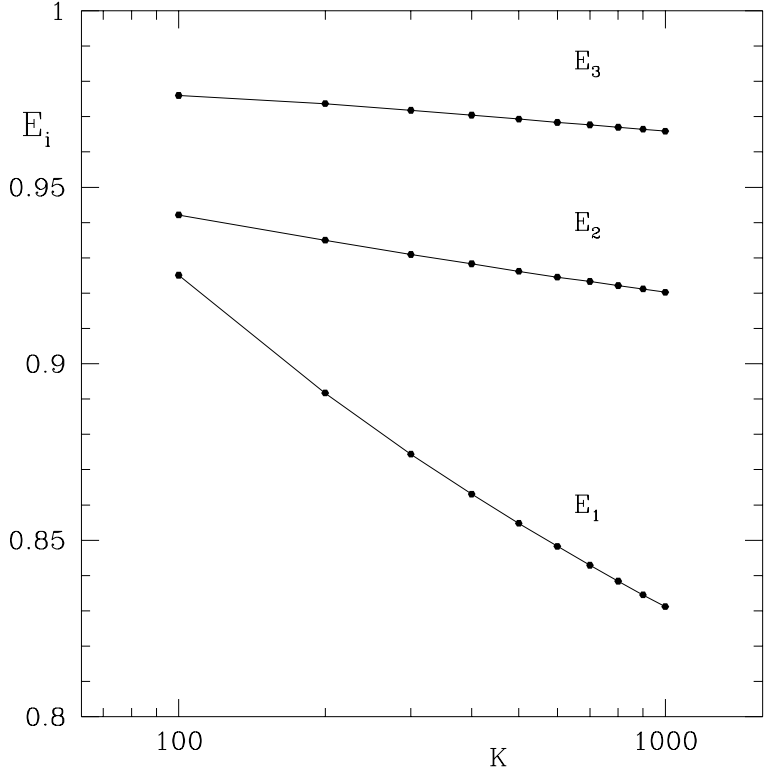,height=11cm,rheight=8cm,angle=-0}
}\vspace*{3cm}
\centerline{ \parbox{35em}{
\caption{ \label {spectrumconvergence} {\sl
 Eigenvalues $E_i$ 
 {\it versus} Large Harmonic Resolution $K$.~-- 
 The lowest three eigenvalues are plotted in however  different 
 mass units $ E_i = P_i^-P_i^+/ m_i^2$ with
 $ m_1^2 =  75 m_u^2$,
 $ m_2^2 = 480 m_u^2$,
 $ m_3^2 = 730 m_u^2$.
 They are reasonably but not completely stable and vary on a level
 of less than 10 per cent. They seem to converge in the continuum
 limit $K\rightarrow\infty$.
}}}}
\end{figure}
Plotting the results in Figure~\ref {spectrumconvergence} on a 
logarithmic scale one is able to unravel logarithmically small 
variations which skip the observation on a linear scale. 
The unexpectedly large but still modest variation of the 
eigenvalues with increasing resolution 
particularly their positive curvature seems to indicate 
convergence to the continuum limit, in line with the analytic 
considerations below in Section~\ref {sec:integral_equation}.
We have verified this behaviour numerically also for 
larger values of the mass parameter $\mu$, but renounce to 
display the results. For negative values of the mass parameter,
however, {\it i.e.} for $\mu^2 < 0$ the eigenvalue curves
tend to develop an increasingly strong negative curvature
with respect to $\ln K$, again in line with analytic considerations. 

\section{Coupled Integral Equations in the Continuum Limit}
\label {sec:integral_equation}

According to Eq.(\ref {def:space}), one deals with orthogonal vector 
spaces $ \left\langle m \vert n \right\rangle = 0 $. In such a space 
the eigenvalue problem 
$ P^- \left| {\Psi_i}  \right\rangle 
= P^- _i \left| {\Psi_i}  \right\rangle $ 
becomes a set of two coupled matrix equations, {\it i.e.}
\begin {eqnarray}
    \sum _{m ^\prime = {1\over 2}} ^{K-{1\over2}}
    \left\langle {n} \right| P ^- \left| {m ^\prime} \right\rangle 
    \left\langle m ^\prime \vert \Psi_i\right\rangle 
&=& \left( P ^- _i - \left\langle {n} \right| P ^- 
    \left| {n}  \right\rangle \right) 
    \left\langle n \vert \Psi_i\right\rangle 
\ , \\
    \sum _{n ^\prime = 1} ^{[K/2]} 
    \left\langle {m} \right| P ^- \left| {n ^\prime} \right\rangle 
    \left\langle n ^\prime \vert \Psi_i\right\rangle 
  + \ \ -\!\!\!\!\!\!\!\!\sum 
    _{m ^\prime = {1\over 2}} ^{K-{1\over2}}
    \left\langle {m} \right| P ^- \left| {m ^\prime} \right\rangle 
    \left\langle m ^\prime \vert \Psi_i\right\rangle 
&=& \left( P ^- _i - \left\langle {m} \right| P ^- 
    \left| {m}  \right\rangle \right) 
    \left\langle m \vert \Psi_i\right\rangle 
\ , \label {eq:coupled_matrix} \end {eqnarray}
see also Eq.(\ref {def:eigenfunctions}).
As mentioned, matrix elements like 
$ \left\langle {n} \right| P ^- \left| {n ^\prime} \right\rangle $ 
vanish in the present model. 
Hence forward we substitute $ P ^- _i  = \omega _i/ P ^+ _i$
and drop the index $i$. 
The notation accounts for the obvious fact that 
the off-diagonal matrix elements are calculated differently 
from the diagonal ones and that the latter are taken to the r.h.s.
of the equations. 
Correspondingly, in the continuum limit, one is confronted with
a set of two {\it coupled integral equations} 
\begin {eqnarray}
    \omega\, \psi _a (x)
&=& I _a (x)\, \psi _a (x) +
    -\!\!\!\!\!\!\int _0 ^1 \!\!dx^\prime\,
    \left\langle {x} \right| K _a 
    \left| {x ^\prime}  \right\rangle \psi _b (x ^\prime)
\ , \label {eq:coupled_integral_a} \\
    \omega\, \psi _b (x)
&=& I _b (x)\, \psi _b (x) +
    -\!\!\!\!\!\!\int _0 ^1 \!\!dx^\prime\,
    \left\langle {x} \right| K _b 
    \left| {x ^\prime}  \right\rangle \psi _b (x ^\prime)
  + -\!\!\!\!\!\!\int _0 ^{1\over2} \!\!dx^\prime\,
    \left\langle {x} \right| K _a 
    \left| {x ^\prime}  \right\rangle \psi _a (x ^\prime)
\ . \label {eq:coupled_integral_b} \end {eqnarray}
More specifically, the continuum limit is obtained 
by the limiting procedure 
$ n \rightarrow \infty $ and $ K \rightarrow \infty $
at {\it fixed momentum fractions} $x = n/K $, 
replacing sums by integrals like for example 
\begin {equation}
   \ \ -\!\!\!\!\!\!\!\!\!\sum 
   _{m ^\prime = {1\over 2}} ^{K-{1\over2}} F (m, m ^\prime)
 = -\!\!\!\!\!\!\int _0 ^1 
   \!\! dx^\prime\,F (x , x^\prime)
   \ , \qquad{\rm with} \ F (x,x^\prime)\equiv K \, F (m, m ^\prime) 
\ . \end {equation}
Care has to be taken of a proper removal of the 
`diagonal kernels', {\it i.e.} 
\begin {equation}
    -\!\!\!\!\!\!\int _0 ^1 \!\!dx^\prime\, F (x , x^\prime)
  \equiv
    \int _0 ^{x - \epsilon} \!\!dx^\prime\, F (x , x^\prime)
  + \int _{x + \epsilon} ^1 \!\!dx^\prime\, F (x , x^\prime)
\ , \end {equation}
with the limit $\epsilon \rightarrow 0$ taken at the end.
Matrix elements are thus related to kernels by
\begin {equation}
    \left\langle {x} \right| K _a \left| {x ^\prime} \right\rangle 
   = K P ^+ {\pi \over L} 
    \left\langle {n} \right| P ^- 
    \left| {m ^\prime} \right\rangle 
\ , \quad{\rm and}\qquad
    \left\langle {x} \right| K _b \left| {x ^\prime} \right\rangle 
   = K \, P ^+ {\pi \over L} 
    \left\langle {m} \right| P ^- 
    \left| {m ^\prime} \right\rangle 
\ . \end {equation}
From Table~\ref {tab:seagulls} one gets 
\begin {eqnarray}
    \left\langle {x} \right| K _a \left| {x ^\prime}  \right\rangle 
&=& - {2 \hat g ^2 \over \left( x - x ^\prime \right) ^2}
      { \left( x+x^\prime \right) \left( 2-x-x^\prime \right) 
        \over \sqrt{x(1-x)\, x^\prime(1-x^\prime)} }
    - {2 \hat g ^2 \over \left( 1 - x - x ^\prime \right) ^2}
      { \left( 1+x-x^\prime \right) \left( 1-x+x^\prime \right) 
        \over \sqrt{x(1-x)\, x^\prime(1-x^\prime)} }
\ , \label {eq:kernel_a} \\ 
    \left\langle {x} \right| K _b \left| {x ^\prime}  \right\rangle 
&=& - {2 \hat g ^2 \over \left( x - x ^\prime \right) ^2}
      { \left( x+x^\prime \right) \left( 2-x-x^\prime \right) 
        \over \sqrt{x(1-x)\, x^\prime(1-x^\prime)} }
    + 2 \hat g ^2 
      { \left( 1-2x \right) \left( 1-2x ^\prime \right) 
        \over \sqrt{x(1-x)\, x^\prime(1-x^\prime)} }
\ . \label {eq:kernel_b} \end {eqnarray}
In the sequel the obvious symmetry
$    \left\langle {x} \right| K _a 
     \left| {x ^\prime}  \right\rangle 
   = \left\langle {x} \right| K _a \left| {1 - x ^\prime}
 \right\rangle $  
will be used without further mentioning. Note the
non-integrable quadratic singularities at
$ x ^\prime = x $ and $ x ^\prime = 1 - x $.
The wavefunctions become 
$ \left\langle m \vert \Psi\right\rangle = \psi _a (x) $ and  
$ \left\langle n \vert \Psi\right\rangle = \psi _b (x) $.
The diagonal terms are obtained from 
Table~\ref {tab:contractions}, {\it i.e.}
\begin {equation} 
    I _a (x) = { C \hat g ^2 \over x(1-x)} + I _b(x) 
\ , \quad{\rm and}\quad 
    I _b (x) = {\mu^2 + 16 \hat g ^2 \over x(1-x)} 
  + -\!\!\!\!\!\!\int _0^1 
    \!\! { 16 \hat g ^2 \,dz \over (x-z) ^2 } 
\ . \label {eq:diagonal} \end {equation} 
We now seek the solution in the regime which corresponds to 
the color singlets, namely 
\begin {equation}
   \psi _b (x) = \psi _b (1-x) = \psi _a (x) \equiv \psi(x)
\ . \end {equation}
The coupled integral equations (\ref {eq:coupled_integral_a})
and (\ref {eq:coupled_integral_b}) degenerate then 
into two equations for one function
\begin {eqnarray}
    \omega\, \psi (x)
&=& I _a (x)\, \psi (x) +
    -\!\!\!\!\!\!\int _0 ^1 \!\!dx^\prime\,
    \left\langle {x} \right| K _a 
    \left| {x ^\prime}  \right\rangle \psi (x ^\prime)
\ , \qquad\qquad{\rm and} \label {eq:integral_a} \\
    \omega\, \psi (x)
&=& I _b (x)\, \psi (x) +
    -\!\!\!\!\!\!\int _0 ^1 \!\!dx^\prime\,
    \left\langle {x} \right| K _b 
    \left| {x ^\prime}  \right\rangle \psi (x ^\prime)
  + {1\over 2} -\!\!\!\!\!\!\int _0 ^1 \!\!dx^\prime\,
    \left\langle {x} \right| K _a 
    \left| {x ^\prime}  \right\rangle \psi (x ^\prime)
\ . \label {eq:integral_b} \end {eqnarray}
One of them must therefore be redundant. 
To solve a singular equation like (\ref {eq:integral_a}) 
one adds and subtracts a term like
$ I(x) = \int _0 ^1 dx^\prime\,
    \left\langle {x} \right| K _a 
    \left| {x ^\prime}  \right\rangle $, {\it i.e.} 
\begin {equation} 
    \omega\, \psi (x)
  = \bigl( I (x) + I _a (x) \bigr)\ \psi (x) +
    -\!\!\!\!\!\!\int _0 ^1 \!\!dx^\prime\,
    \left\langle {x} \right| K _a 
    \left| {x ^\prime}  \right\rangle 
    \ \bigl( \psi (x ^\prime) - \psi (x) \bigr) 
\ . \label {eq:integral_2} \end {equation} 
Close to the singularity, 
$ \psi (x) - \psi (x^\prime ) \sim x - x^\prime $
converts the quadratic singularity into a pole which
can be integrated by some principal value prescription.
Now look at 
\begin {equation} 
    I (x) + I _a (x) = {\mu ^2 + 16\hat g ^2 + C\hat g ^2 \over
      x(1-x)} 
  + 4\hat g ^2 -\!\!\!\!\!\!\int _0^1 \!\! 
    { dx^\prime \over (x-x^\prime) ^2 } 
    \left[ 4 - {\left( x+x^\prime \right) \left( 2-x-x^\prime \right)  
                \over \sqrt{x(1-x)}
                \sqrt{x^\prime(1-x^\prime)}}\right] 
\ . \end {equation} 
Setting in the square bracket $ x ^\prime = x + \Delta $ and 
expanding with $ \Delta $ one notes that the first 
non-vanishing term is $ \propto \Delta^2 $. This cancels the
singularity, and the integral becomes finite. 
The integral equation (\ref {eq:integral_a}) 
has thus a solution.~-- As for Eq.(\ref {eq:integral_b}) 
we subtract the latter from the former to get
\begin {equation} 
    {C \hat g ^2  \over x(1-x) }\,\psi (x)
  = -\!\!\!\!\!\!\int _0 ^1 \!\!dx^\prime\, \psi(x ^\prime) \left[ 
    \left\langle {x} \right| K _b 
    \left| {x ^\prime}  \right\rangle 
  - {1\over 2} \left\langle {x} \right| K _a 
    \left| {x ^\prime}  \right\rangle \right] 
\ . \label {eq:integral_c} \end {equation} 
Inserting the kernels from Eqs. (\ref {eq:kernel_a})
and (\ref {eq:kernel_b}) the singular terms cancel precisely.
One remains with
\begin {equation}
    {C \hat g ^2 \over x(1-x) }\,\psi (x)
  = 2 \hat g ^2 
    {\left( 1-2x \right) \over \sqrt{x(1-x)}}
    -\!\!\!\!\!\!\int _0 ^1 \!\!dx^\prime\, \psi(x ^\prime) 
    {\left( 1-2x ^\prime \right) \over \sqrt{x^\prime(1-x^\prime)}}
  = 0
\ , \label {eq:integral_z} \end {equation} 
because the integrand is an odd function 
around $x ^\prime = {1\over2}$. 
One has to conclude:
The second integral equation is consistent with the first one
iff $ C = 0 $.
Otherwise the continuum limit does not exist.
This is in line also with the statement from color invariance:
All three particles should have the same mass, 
see Eq.(\ref {eq:diagonal}).~-- 
One remains with Eq.(\ref {eq:integral_a}) which units adjusted
is identical with the one of Bardeen {\it et al.} \cite {BPR80} 
and of Klebanov {\it et al.} \cite {DKB93}, although the latter 
had been derived in another, {\it i.e.} in the light-cone gauge.
We refer to their work particularly with respect to the endpoint
analysis of the solutions. 

\section{Charge Conservation and Color Singlets}
\label {sec:charges}

We take the multiplet structure of the spectra in 
Section~\ref {sec:numerical_solution} as kind of an empirical fact, 
and wonder whether they can be put on more solid grounds.
This turns out more difficult than anticipated.~-- Of course, 
one always can associate charges with the 
currents as defined in Section~\ref {sec:numerical_solution}, 
particularly $ Q _- = \int dx ^- J ^+_- $, $ Q _+ = Q _- ^\dagger$,
and the familiar $ Q _3$. Upon evaluation they become 
\begin {equation} 
   Q _3 = \sum _{m = {1\over 2}} ^{\infty} 
          d _ m ^\dagger d _ m - b _ m ^\dagger b _ m 
   \,,\quad {\rm and}\quad
   Q _-  = A _0 \Bigl( a _0 b _{{1\over 2}} \Bigr) _{\!\! s} 
         + \sum _{n=1} ^{\infty} 
         A _n  a _n ^\dagger b _{n + {1\over 2}} -
         B _n  a _n d _{n - {1\over 2}} ^\dagger 
\,, \label {eq:charges} \end {equation} 
with 
$ ( a _0 b _{{1\over 2}} ) _s \equiv
  {1\over 2} ( a _0 b _{{1\over 2}} + b _{{1\over 2}} a _0 ) $.
In the fundamental modular domain the coefficients are 
\begin {equation} 
   A _0 = {1\over 2} \sqrt {z} \,, \quad
   A _n = {1\over 2} 
   \left( \sqrt {{ n+z\over n}} + \sqrt {{ n\over n+z}} \right), 
   \quad {\rm and}\quad
   B _n = {1\over 2} 
   \left( \sqrt {{ n-z\over n}} + \sqrt {{ n\over n-z}} \right)
\,. \label {app-eq:coefficients_app} \end {equation} 
Since $\Bigl[ a _0\,,\, Q _3 \Bigr] = 0$, see Ref.\cite {Kal95}, 
one calculates the SU(2)-commutators as follows: 
\begin {equation} 
\biggl[ Q _3 \ , \ Q _- \biggr] = Q _- 
\quad {\rm and}\quad
\biggl[ Q _- \ , \ Q _+ \biggr] - Q _3 = \Delta(z,a_0)
\,. \label {eq:commutator} \end {equation} 
The right hand side of this equation turns out to be
\begin {eqnarray} 
\lefteqn{   {4\over z} \Delta(z,a_0) \equiv 
   \biggl[ \Bigl( a _0 b _{{1\over 2}}          \Bigr) _{\!\! s} \ , \
           \Bigl( a _0 b _{{1\over 2}} ^\dagger \Bigr) _{\!\! s}
           \biggr]  
 + }\nonumber \\ 
   &\hspace*{-3.5em}+\hspace*{-2.5em}&\hspace{-0.8em}   {2 \over \sqrt
     {z}}  
   \biggl[ \Bigl( a _0 b _{{1\over 2}} \Bigr) _{\!\! s} \ , \ %
   \sum _{n=1} ^{\infty} 
      A _n  a _n ^\dagger b _{n + {1\over 2}} 
   -  B _n  a _n d _{n - {1\over 2}} ^\dagger          \biggr] 
   -   {2 \over \sqrt {z}} 
   \biggl[ \Bigl( a _0 b _{{1\over 2}} ^\dagger \Bigr) _{\!\! s} \ , \ 
   \sum _{n=1} ^{\infty} 
      A _n  a _n b _{n + {1\over 2}} ^\dagger 
   -  B _n  a _n ^\dagger d _{n - {1\over 2}}          \biggr] 
 \nonumber \\ 
   &\hspace*{-3.5em}+\hspace*{-2.5em}&\hspace{-0.8em}   \sum _{n=1}
   ^{\infty} { z \over n ^2 - z ^2} 
   \left( d _{n - {1\over 2}} ^\dagger d _{n - {1\over 2}} 
        - b _{n + {1\over 2}} ^\dagger b _{n + {1\over 2}}    \right)
   +   \sum _{n=1} ^{\infty} 
       { z ^2 \over n\left(n ^2 - z ^2\right)}
   \left( d _{n - {1\over 2}} ^\dagger d _{n - {1\over 2}} 
        + b _{n + {1\over 2}} ^\dagger b _{n + {1\over 2}} 
        + a _n ^\dagger a _n                                  \right)
   \,. \label {eq:delta} \end {eqnarray} 
In passing one notes that the charges of Eq.(\ref {def:Q_+}) 
agree with those of Eq.(\ref {eq:charges}) 
for $z=0$ (thus $\Delta = 0$), and for $a _0 = 0$.
Can one anticipate for the general case that
$ \Delta ( z, a _0) = 0 $?~-- One has two perspectives for the
future: Either 
(I)  the explicit solution for $a_0$ renders $\Delta=0$, or 
(II) one requires $\Delta=0$ as a subsidiary condition for
solving $a_0$. Here the matter rests as long as one cannot solve 
explicitly for $ a _0$. 

But in either case seems to be a problem: 
The charges as defined by Eq.(\ref {eq:charges}) cannot be
conserved since $ \partial _\beta {\bf J} ^\beta \neq 0 $.
True, our model has a conserved four-current 
$ \partial _\beta \widetilde {\bf J} ^\beta = 0 $, since this
follows directly from the color-Maxwell equations
$ \partial _\alpha {\bf F}^{\alpha\beta} 
  = g \widetilde {\bf J} ^\beta $. 
But the two currents are not identical since
$   \widetilde {\bf J} ^\beta = {\bf J} ^\beta + {1\over i} 
    \bigl[ {\bf F} ^{\beta\alpha} \,,\, {\bf A} _\alpha \bigr] $.
Only for the latter holds
\begin {equation} 
   {d\over dx^+} \widetilde {\bf Q} = 0 
   \,, \qquad{\rm with}\quad 
   \widetilde {\bf Q} 
   = \int \limits _{-L} ^{+L} dx ^- \widetilde {\bf J} ^+ 
\,. \label {app-eq:cons-charge} \end {equation} 
Thus 
$ \widetilde {\bf Q} = {\bf Q} + 2 i L
     [ \partial _+ {\bf A} ^+ \,,\, {\bf A} ^+ ] 
  +  2Lg \bigl[ [ {\bf A} ^+ \,,\, 
       \left\langle{\bf A} ^-\right\rangle_{\! 0} ]\,,\,
       {\bf A} ^+ \bigr] $, or
\begin {equation}
    \widetilde Q _3  
  = Q _3 
\,, \quad{\rm and}\qquad   
    \widetilde Q _-  = Q _-  
  -  2L g v^2 \bigl\langle A ^- _+ \bigr\rangle_{\! 0} 
\,. \label {eq:charges-explicit} \end {equation}
But now, in the process of inverting the Gauss' 
equations (\ref {gausscomp}), their zero modes have been required 
to vanish \cite {PKP95}. This in turn leads to
$ \widetilde Q _\pm \equiv 0 $, in total and unpleasant opposition 
to tribal beliefs that the true and dynamically conserved charges 
$\widetilde Q _a$ obey a non-Abelian group structure. 

As a possible way out we proffer a different gauge choice namely
\begin {equation} 
    \left< {\bf A} ^- \right> _{0} = 0
\,, \label {eq:gauge_complem} \end {equation} 
{\it in addition to} Eq.(\ref {eq:gauge}). We leave it for future 
work to check whether this choice is possibly in conflict with the 
first gauge condition in Eq.(\ref{eq:gauge}). 
But the proffered choice would have the advantage 
to be manifestly invariant under color rotations and to treat 
the zero mode components of ${\bf A} ^-$ the same way in all 
three Gauss' equations(\ref {gausscomp}). Moreover,
since $\widetilde {\bf Q} = {\bf Q} $, both charges would be 
strict constants of the motion, reconciling thus naive and 
refined considerations.

But perhaps these considerations are academic 
since all zero modes become sets of measure zero in the continuum 
limit. The solutions of the integral equation, for example, 
are color singlets in the strict sense. As mentioned, 
they are independent of the gauge choices in the zero mode sector.

\section{Summary and Discussion}
\label {sec:summary}

The approach taken here follows closely some earlier work \cite{PKP95}.
Beginning with SU(2) pure gauge theory in (2+1) dimensions in the
front form one suppresses the transverse coordinate dependence 
of the gluons and obtains a (1+1) dimensional gauge theory
coupled to adjoint scalar matter. 
The present work has one major aim: 
We would like to get a first and rough idea on 
how the excitation spectrum as well as 
how the structure functions look 
in such a dimensionally reduced quantum field theory.
Therefore, in order to have a tractable formalism and in contrast 
to earlier \cite{PKP95} and ongoing work \cite{Kal95}, we suppress 
here by hand the dynamical impact of the topological gauge zero mode 
($\zeta$) and of the constrained zero mode ($a_0$). 
We perform furthermore a Tamm-Dancoff truncation and include only
the 2-particle Fock space. 

What are the results? Most important we think is the result
displayed  in Figure~\ref {separate}, namely that the solutions 
corresponding to color singlets ($Q=0$) have a discrete and finite
spectrum over a wide range of the harmonic resolution $K$. 
The states corresponding to color triplets ($Q=1$) or pentuplets 
($Q=2$) can well be separated and tend to have a very large 
if not infinite mass. This appears to be one of the few available
concrete pieces of evidences that only the color singlets can 
have finite mass in a non-abelian field theory. 
The method to identify these states is somewhat pragmatic since 
the model assumption on the zero mode, $a_0=0$, violates strict 
SU(2) invariance. 

In Section~\ref {sec:integral_equation} the continuum limit was
established by deriving an integral equation in the manner of 
Bergknoff \cite{ber77} for QED$_{1+1}$, see also \cite{epb87}. 
As demonstrated, the continuum limit exists only for $C=0$,
resulting for the color singlets in the integral equation 
\begin {equation} 
    \omega\, \psi (x)
    = {\mu ^2 + 16\hat g ^2 \over x(1-x)} \,\psi (x)
  + 4\hat g ^2 -\!\!\!\!\!\!\int _0^1 \!\! 
    { dx^\prime \over (x-x^\prime) ^2 } 
    \left[ 4\,\psi (x) - 
    {\left( x+x^\prime \right) \left( 2-x-x^\prime \right) 
                \over \sqrt{x(1-x) \,x^\prime(1-x^\prime)}}
    \, \psi (x ^\prime) \right]
\,. \label {eq:final_int} \end {equation} 
The integral equation is identical with the one derived earlier 
by Bardeen {\it et al.} \cite {BPR80} and by 
Klebanov {\it et al.} \cite {DKB93}, a remarkable fact in view
that both these authors have used the light-cone gauge as opposed
to the present light-cone {\it Coulomb} gauge.

The present model is restricted to the lowest non-trivial
Fock states, those with particle number two. 
Fock states with three partons like 
$a_n^\dagger b_m^\dagger d_k^\dagger \left| 0 \right\rangle $ 
are ruled out since all matrix elements between 
the 2- and the 3-particle sector vanish due to color invariance.
The next higher Fock-states have thus four partons: aaaa-,
aabd, and bdbd-states. Based on the numerical experience with
the massive and massless Schwinger model \cite {epb87,epk95}, 
one can guess that the admixtures of the 4-parton states
in the low lying part of the spectrum are of the order of $10^{-3}$
and thus probably negligible. 

A rather interesting feature of the model should be mentioned. 
For sufficiently small variations of the mass parameter $\mu$ 
around the value zero the spectrum as displayed in 
Figure~\ref {spectrum1} can be moved up and down unchanged almost 
at gusto, including the case that the lowest eigenvalue coincides 
with zero. A trace of that survives even in the larger 
variations of Figure~\ref {mudependence1}. 
It looks {\it as if} a mass {\it difference} like the one in 
Eq.(\ref {eq:masses}) is a dimensionless number characteristic
for the present model. It should be interesting to get an 
analytic estimate, because the mechanism allows to generate 
{\it huge mass ratios} $M_2/M_1$ as they are characteristic for 
hadronic physics.

What can one learn from this work beyond our particular play model?
Let us return to the fundamental assumption 
$  \partial_i  {\bf A}^\mu = 0 $ as in \cite{PKP95}.
Dimensionful quantities are never strictly zero, so let us 
be more quantitative. In the full 3+1 dimensional treatment one would 
introduce an invariant mass cut-off like for example 
the one of Lepage and Brodsky, see \cite {PaB91}, 
\begin {equation}
   \sum _\nu \left( {m^2 + \vec k _\perp ^2 \over x(1-x) } \right)
   _\nu 
   \leq \Lambda ^2
\ . \end {equation}
It is covariant, the sum runs over all partons $\nu$. 
The scale $\Lambda$ is at our disposal. For two massless gluons one 
obtains thus $ \vec k _\perp ^2 \leq x(1-x) \Lambda ^2$. 
Maximizing this by $x=1/2$ and using transverse periodic 
boundary conditions with $ \vec k = \vec n \pi / L_\perp$ one gets 
$ \vert n \vert \leq \Lambda L _\perp /(2 \pi) $.
Therefore, all transverse momenta are cut out when one {\it defines} 
the transverse length by
\begin {equation}
    L _\perp \equiv {\pi \over \Lambda }
\ . \label {eq:length} \end {equation}
Only the transverse zero modes
survive, and this is precisely the present model. 

The fact that we have started from 2+1 dimensions weighs less. 
In 3+1 dimensions, the model with only transversal zero modes
in the manner of \cite{PKP95} is only marginally more
complicated due to the non-abelian commutator term. 
First estimates show that its impact is not dominant. 
With a grain of salt, the present color singlet solutions 
would then correspond to the helicity aligned 
$ 2 ^{++}$-glue-balls.
If one sets the cut-off scale as the typical hadronic energy 
$\Lambda \sim 1\ GeV$ one gets for the transversal size 
$ L_\perp \sim 0.6\ fm$, a value not untypical 
for hadronic sizes. 

One can push these simple considerations even further.
The present coupling constant $g$ is related to $g_3$ in 
3+1 dimensions \cite{PKP95} 
by $g_3 = 2 g L_\perp $. 
Our unit of mass $m _u = g /4 \sqrt\pi$ can thus be expressed
in terms of $\alpha _s = g_3 ^2/ 4\pi$ and $\Lambda$. 
A value of $\alpha_s \sim 0.4$ looks reasonable as compared
to the empirical $\alpha_s(M_Z) \simeq 0.12$.
With Eq.(\ref {eq:length}) one obtains 
\begin {equation}
   m_u = \sqrt {\alpha_s} {\Lambda \over 8\pi } \sim 25\ MeV
\ , \end {equation}
which allows to convert our numerical results to physical units. 
By order of magnitude, one reads off Figure~\ref {spectrum1} 
for the first two states $M_1^2 \simeq 0.0018\,m_0^2 $ and 
$M_2^2\simeq 0.038\,m_0^2 $, thus 
\begin {equation}
    {1\over m_u^2} \left( M_2^2 - M_1^2 \right) \simeq 800
\ , \qquad 
    M_1 \simeq 150\ MeV
\ , \qquad 
    M_2 \simeq 600\ MeV  
\ . \label {eq:masses} \end {equation}
These numbers are not untypical for hadrons.

In conclusion, the physical picture emerging from this work
might over-stress the point but without being necessarily false:
Hadrons particularly glue-balls have no transversal structure
up to transversal sizes of about $ 0.5 \ fm$, compare also with the 
recent work of van Baal \cite {vBa92}. The field lines are 
parallel due to the basic ingredient of the model, 
in accord with phenomenological flux-tube models. Up to 
these sizes, the structure resides in the longitudinal 
direction as mirrored in the
structure functions displayed in Figure~\ref {structure1}. 

\section{Acknowledgement}
The authors thank Dr. Alex Kalloniatis for pursuing 
the progress of this work which would not have started without him. 
We have enjoyed the many discussions, but regret that we might have 
distracted him from his perennial fight with the constraint equation.
We thank Dr. Brett van de Sande for suggesting mass renormalization
and  appreciate the comments of Drs. M.~Burkardt and and S.~Dalley
during their visits at the Max-Planck Institut in the summer 1995.

\begin {appendix}
\newpage 
\begin {appendix}
\section{The Matrix Elements of the Contraction Energy}
\label {sec:contractions}

\begin{table}[H]
\centerline{ \parbox{35em}{
\caption{ \label {tab:contractions} {\sl
The contraction part $P^-_C$ of the Hamiltonian is expressed
in terms of Fock-space operators.
The coupling constant is absorbed into the
coefficient $\displaystyle \hat g ^2=\frac{g^2}{16\pi}$. 
As discussed in the text, one should use $C=0$ in the calculations.
}}}}
\vspace*{-2.5cm}
\centerline{ \psfig{figure=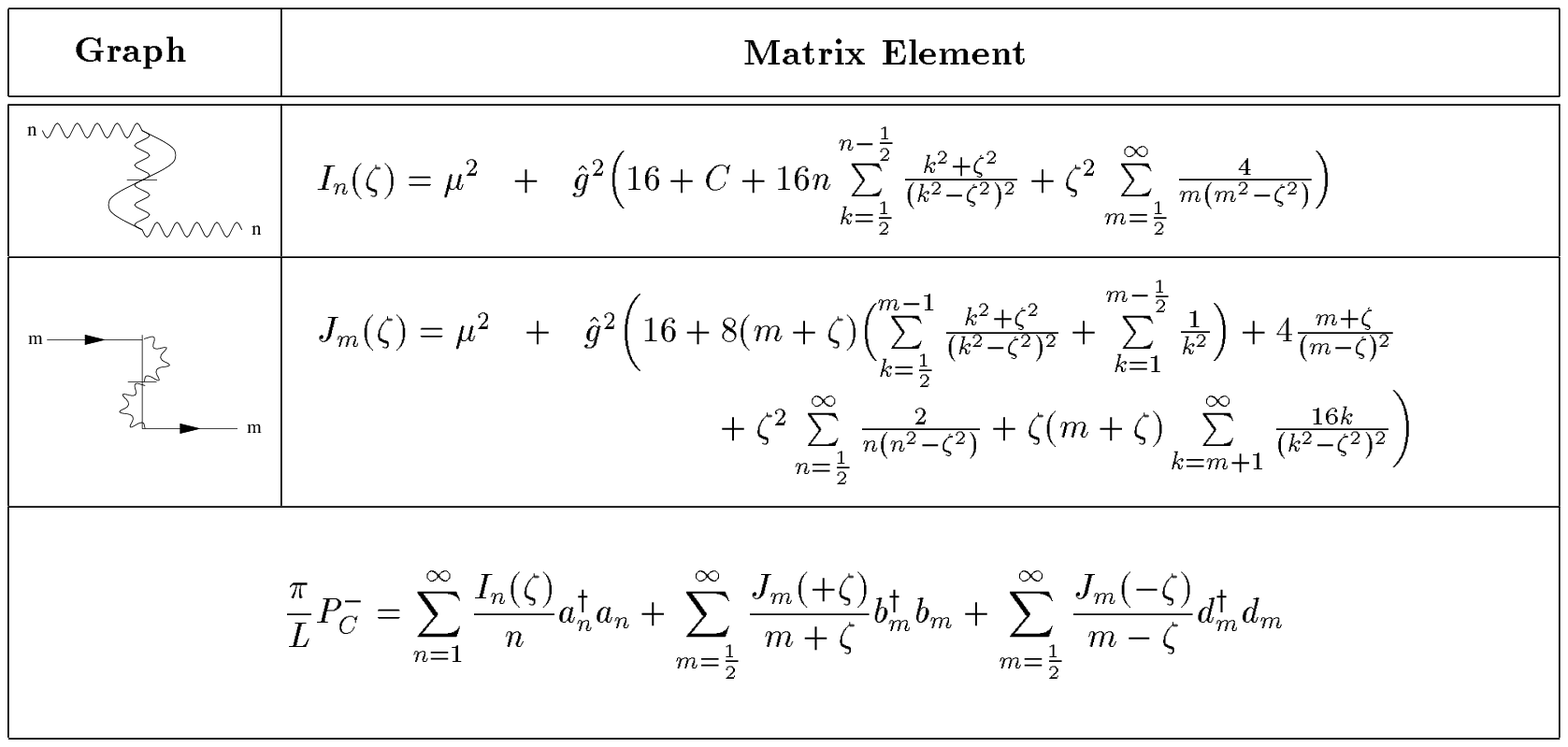,width=21cm,rheight=2cm,clip=,angle=0}}
\end{table}

\newpage 
\section{The Matrix Elements of the Seagull Energy}
\label {sec:seagulls}

\begin{table}[H]
\centerline{ \parbox{35em}{
\caption{ \label {tab:seagulls} {\sl 
The seagull part $P^- _S$ of the Hamiltonian 
is expressed in terms of Fock-space operators.
The coupling constant is absorbed into the coefficient 
$\displaystyle \hat g ^2=\frac{g^2}{16\pi}$. 
The indices of $a$ and $a^\dagger$  are integers, those of
$b,b^\dagger,d$ and $d^\dagger$ half-integers. 
The Kronecker delta $\delta^{n_1+n_2}_{n_3+n_4}$ referring to 
momentum conservation in  $S(n_1,n_2;n_3,n_4)$ is not kept track of
explicitly.
}}}}
\vspace*{-2.5cm}
\centerline{ \psfig{figure=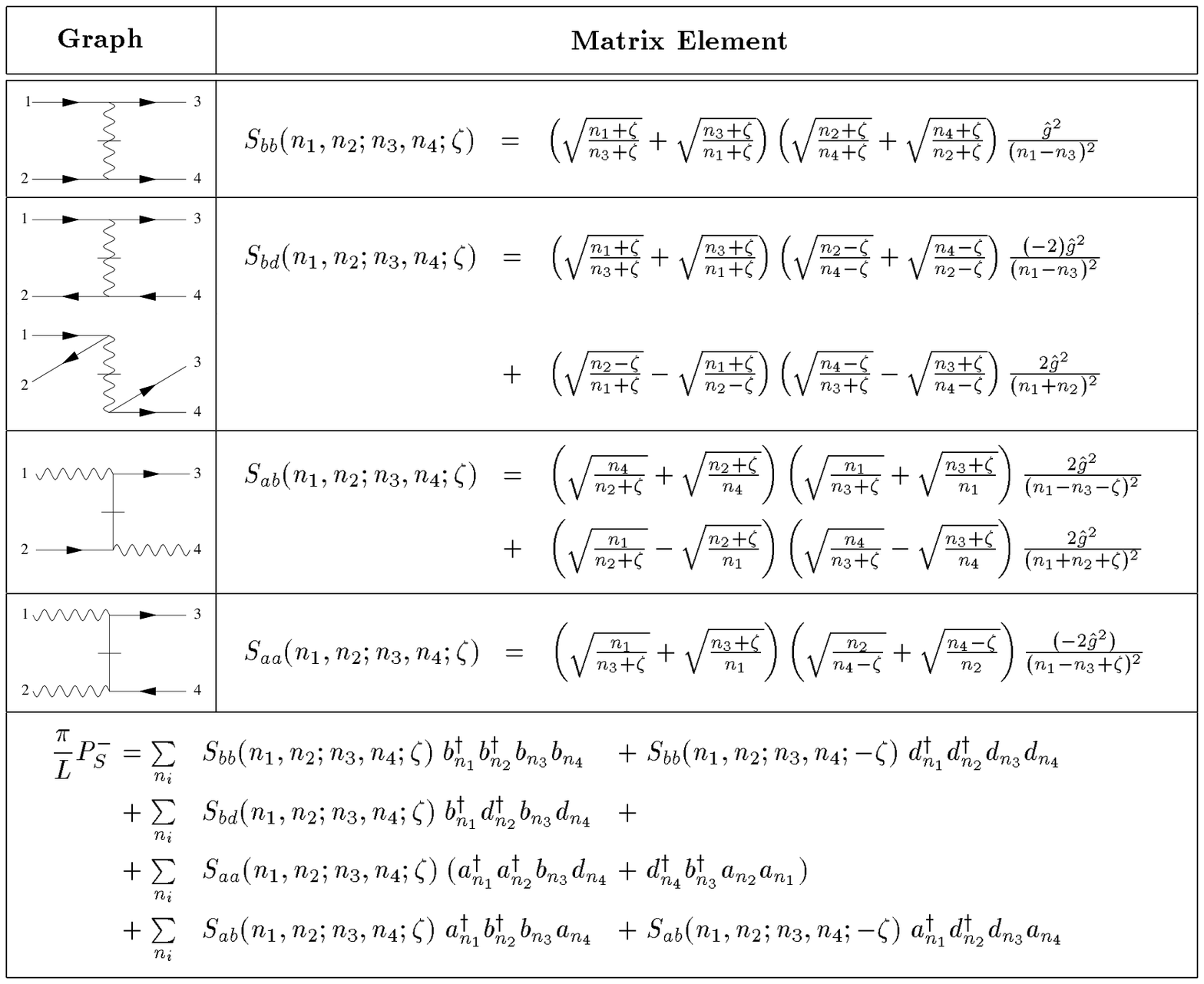,width=21cm,angle=-0}}
\end{table}
 
\newpage 
\section{The Matrix Elements of the Fork Energy}
\label {sec:forks}
\begin{table}[H]
\centerline{ \parbox{35em}{
\caption{ \label {tab:forks} {\sl
The fork part $P^- _F$ of the Hamiltonian
is expressed in terms of Fock-space operators.
The coupling constant is absorbed into the coefficient 
$\displaystyle \hat g ^2=\frac{g^2}{16\pi}$. 
The indices of $a$ and $a^\dagger$  are integers, those of
$b,b^\dagger,d$ and $d^\dagger$ half-integers. 
The Kronecker delta $\delta ^{n_1} _{n_2+n_3+n_4}$ referring to 
momentum conservation in  $F(n_1;n_2,n_3,n_4)$ is not kept track of
explicitly.
}}}}
\vspace*{-2.5cm}
\centerline{ \psfig{figure=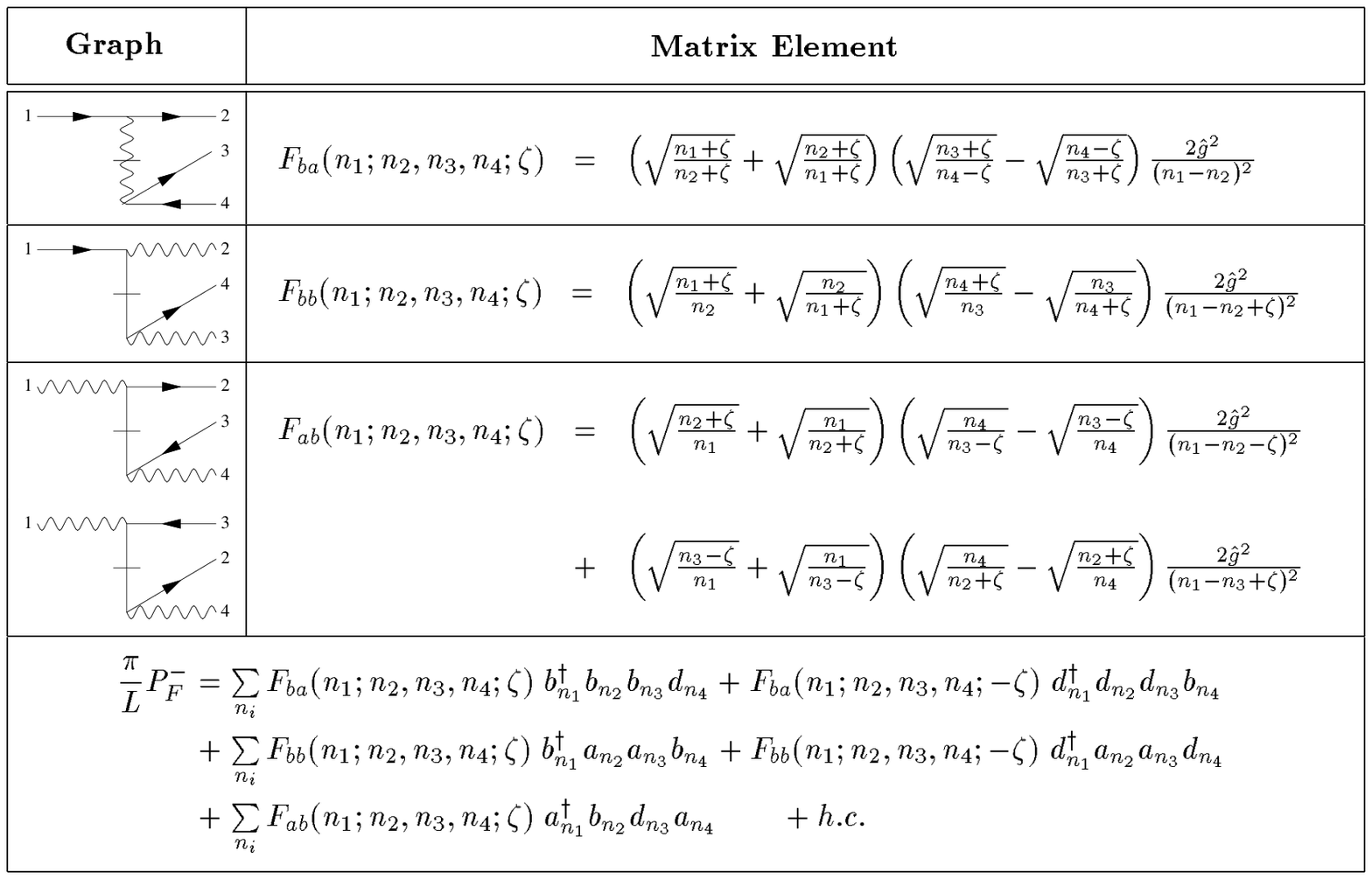,width=21cm,angle=-0}}

\end{table}
\end{appendix}

\end{appendix}
\end   {document}